\documentclass[twocolumn,showpacs,preprintnumbers,aps,prl,amsmath,amssymb,floatfix,superscriptaddress]{revtex4-1}

\usepackage{graphicx}% Include figure files
\usepackage{dcolumn}% Align table columns on decimal point
\usepackage{bm}% bold math
\usepackage{epsfig}
\usepackage{color}
\usepackage{natbib}

\newcommand{\ket}[1]{\left|{#1}\right\rangle}

\begin{document}

\title{Coherent heteronuclear spin dynamics in an ultracold spin-1 mixture}
\author{Xiaoke Li}
\affiliation{Department of Physics, The Chinese University of Hong Kong, Hong Kong, China}
\author{Bing Zhu}
\affiliation{Department of Physics, The Chinese University of Hong Kong, Hong Kong, China}
\author{Xiaodong He}
\affiliation{Department of Physics, The Chinese University of Hong Kong, Hong Kong, China}
\author{Fudong Wang}
\affiliation{Department of Physics, The Chinese University of Hong Kong, Hong Kong, China}
\author{Mingyang Guo}
\affiliation{Department of Physics, The Chinese University of Hong Kong, Hong Kong, China}
\author{Zhi-Fang Xu}
\affiliation{Department of Physics and Astronomy, University of Pittsburgh, Pittsburgh, Pennsylvania 15260, USA}
\author{Shizhong Zhang}
\affiliation{Department of Physics and Centre of Theoretical and Computational Physics, The University of Hong Kong, Hong Kong, China}
\author{Dajun Wang}
\email{djwang@phy.cuhk.edu.hk}
\affiliation{Department of Physics, The Chinese University of Hong Kong, Hong Kong, China}

\date{\today}

\begin{abstract}
We report the observation of coherent heteronuclear spin dynamics driven by inter-species spin-spin interaction in an ultracold spinor mixture, which manifests as periodical and well correlated spin oscillations between two atomic species. In particular, we investigate the magnetic field dependence of the oscillations and find a resonance behavior which depends on {\em both} the linear and quadratic Zeeman effects and the spin-dependent interaction. We also demonstrate a unique knob for controlling the spin dynamics in the spinor mixture with species-dependent vector light shifts. Our finds are in agreement with theoretical simulations without any fitting parameters.

\end{abstract}

\maketitle

Understanding collective spin dynamics is a problem of fundamental importance in modern many-body physics. For example, it underlies the pursuit of spintronics in which spin, rather than charge, is the primary carrier for information processing~\cite{Zutic2014}. Central to the understanding of spin dynamics is the role of spin-spin interactions, which exists between identical as well as distinguishable spins, and their interplay with the (linear and quadratic) Zeeman effects. In this regard, the ultracold spinor quantum gas~\cite{Kawaguchi2012,Kurn2013} provides a powerful platform for investigating spin dynamics due to the high controllability. So far, a rich variety of phenomena have been explored experimentally, including spin oscillations~\cite{Kuwamoto2004,Schmaljohann2004,Chang2004,Chang2005,Kronjager2006,Black2007,Klempt2009,Widera2005} in spinor Bose-Einstein condensate (BEC) and in thermal Bose gas~\cite{Pechkis2013}, as well as various types of spin textures~\cite{Sadler2006,Leslie2009,Choi2012,Eto2014}. Very recently, coherent spin dynamics and giant spin oscillation of a Fermi sea~\cite{Krauser2014} and its relaxation~\cite{Ebling2014} have been investigated.

Until now, however, spin dynamics in ultracold atoms has been explored only in a single atomic species. Here, we realize a system consisting of distinguishable spin-$1$ atoms of $^{87}$Rb and $^{23}$Na, and demonstrate well-controlled and long-lived coherent spin oscillations between them. Our study brings out several unique features of the spinor mixtures. (1) For collisions between two distinguishable spin-1 atoms, there is no exchange symmetry requirement and the interaction takes place over all possible total spin $F$ channels\cite{Ho98,Ohmi98,Law1998,Luo2007,Xu2009,Xu2010,Yu2010}. On the other hand, only even $F$ are allowed for homonuclear collisions. (2) In a single species spinor gas, spin dynamics is governed by the competition between the spin-dependent interaction energy and the quadratic Zeeman shift~\cite{Zhang2005}, while the linear Zeeman shift can be gauged away as magnetization is conserved. In the case of a spinor mixture, both linear and quadratic Zeeman shifts are important. (3) Because the two species have different electronic structures, a differential effective magnetic field can be generated with a spin-dependent ODT. Together with the external magnetic field, this can be used to further control the spin dynamics.

The interaction between two heteronuclear spin-$1$ bosons at a distance $\textbf{r}$ can be written as~\cite{Ho98,Ohmi98},
\begin{equation}
{V_{12}}\left( {\bf{r}} \right) = \left( {\alpha  + \beta {{\bf{f}}_1} \cdot {{\bf{f}}_2} + \gamma {P_0}} \right)\delta \left( {\bf{r}} \right),
\label{eq:interaction}
\end{equation}
where $\alpha  = (g_1 + g_2)/2$ represents the spin-independent interaction. The strength of spin-dependent term is given by $\beta  = (g_2 - g_1)/2$. The third term, $\gamma  =( 2 g_0 - 3 g_1 + g_2)/2$, operates only in the total $F=0$ channel via the projection operator $P_0$. ${\bf{f}}_1$ and ${\bf{f}}_2$ label the hyperfine spin of two different atoms. The coupling constants $g_{\rm F} = 2\pi \hbar ^2 a_{\rm F}/ \mu $ are determined by the s-wave scattering lengths ${a_{\rm F}}$ of the corresponding $F$ channels and the reduced mass $\mu$, where $\hbar$ is the Planck's constant. From previous studies~\cite{Wang2013,BoGao}, we have determined ($\alpha$, $\beta$, $\gamma$) = $2\pi \hbar ^2 a_{\rm B}/\mu \times$ (78.9, -2.5, 0.06), where $a_{\rm B}$ is the Bohr radius. Similar to homonuclear spinor gases of $^{23}$Na and $^{87}$Rb, both $\beta$ and $\gamma$ are much smaller than $\alpha$~\cite{Supp}. The negative $\beta$ indicates a ferromagnetic heteronuclear spin-spin interaction, which tends to align the spins of the two species along the same direction.

Let us consider the collision between a $^{87}$Rb atom in spin state $\ket{m_1}$ and a $^{23}$Na atom in spin state $\ket{m_2}$, which we denote as $\ket{m_1,m_2}$ in the following. Here $m = \pm 1, 0$, corresponding to the three Zeeman sub-levels of the $f$ = 1 hyperfine state. The magnetic energy associated with $\ket{m_1,m_2}$ will be denoted as $E^{m_1,m_2}(B)$. The aforementioned $\beta$ and $\gamma$ terms can support several possible heteronuclear spin changing processes as long as the population of each species and the total magnetization along the magnetic field ${B}$, are conserved. This is in stark contrast to the homonuclear spin-1 case, where only one spin changing process $2\ket{0} \leftrightarrow \ket{1}+\ket{-1}$ is allowed. In this work, we focus on the following heteronuclear spin changing  process,
\begin{align}
\ket{0,-1} \leftrightarrow \ket{-1,0},
\label{eqn_spin_exchange}
\end{align}
which is driven solely by the $\beta$ term~\cite{Xu2012}.

\begin{figure}
\includegraphics[width=0.8 \linewidth]{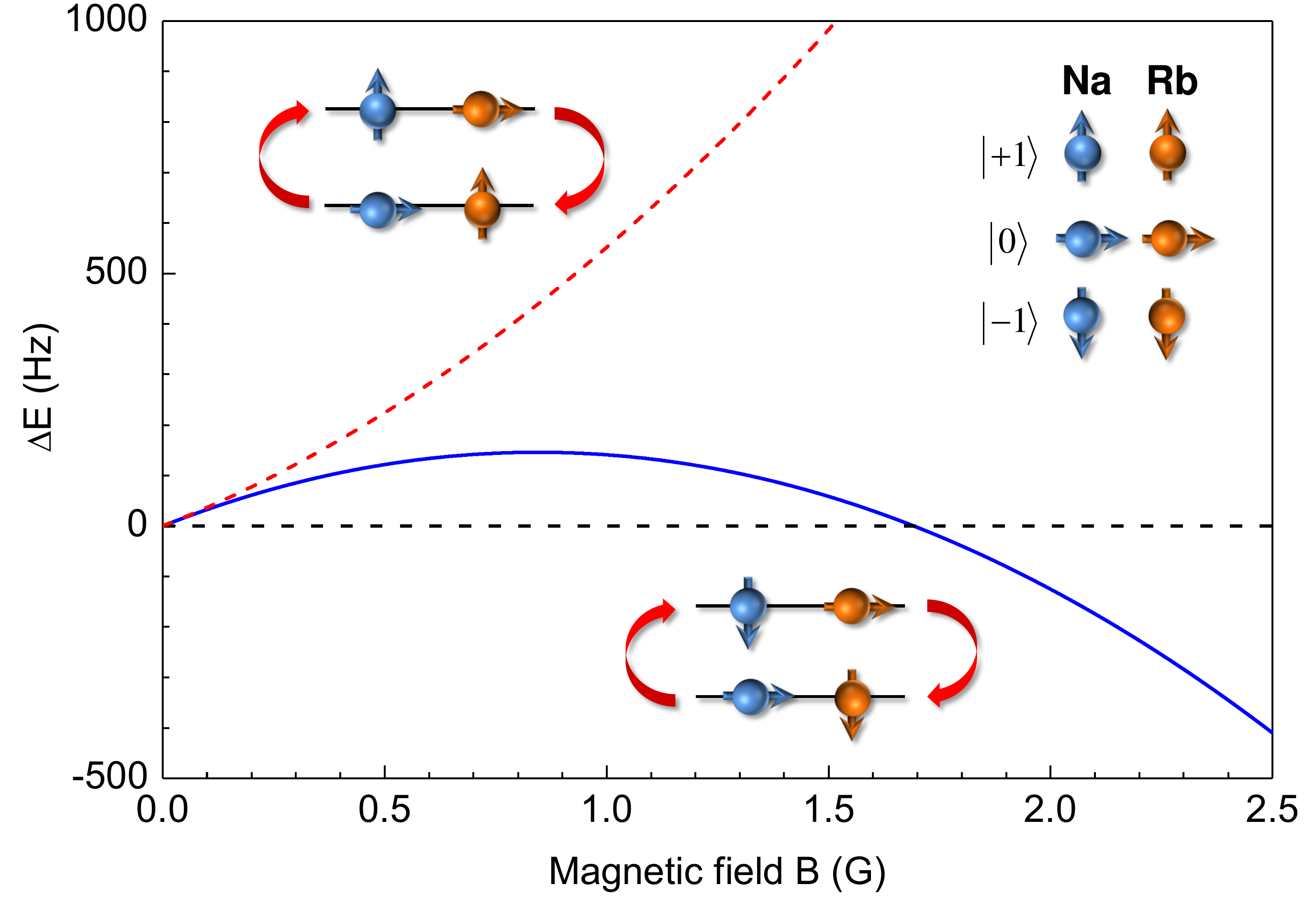}
\caption{\label{fig:principle} (color online) Magnetic energy diagram for two heteronuclear spin changing processes driven by $\beta$ term. The three spin states are represented by three small arrows and atomic species are color coded. $\Delta E$ sets the energy difference between the two states involved. For the process in Eq.~(\ref{eqn_spin_exchange}), $\ket{0,-1}\leftrightarrow \ket{-1,0}$ (blue curve), $\Delta E$ crosses zero at $B_0=1.69$ G, while at the same field, for the process $\ket{0,1}\leftrightarrow \ket{1,0}$ (red curve), $\Delta E$ is around $1000$ Hz. In the vicinity of $B_0$, the small spin-dependent interaction energy couples the two relevant states ($\ket{0,-1}, \ket{-1,0}$) and visible oscillations occur when its strength is comparable to $\Delta E$. While the other homonuclear and heteronuclear processes, including $\ket{0,1}\leftrightarrow \ket{1,0}$, are far detuned and oscillations are greatly suppressed.
}
\end{figure}

Intuitively, coherent spin dynamics of Eq.~(\ref{eqn_spin_exchange}) can be understood from the interplay between the spin-dependent interaction energy ($\beta$ term) and the difference of total Zeeman energy between these two states: $\Delta E(B)\equiv E^{0,-1}(B)-E^{-1,0}(B)$, as depicted in Fig.~\ref{fig:principle}. In analogy to a driven two-level system, when the two energies are very different, the system undergoes detuned oscillations with large frequency but small amplitude. On the other hand, when the two energies are comparable, the system oscillates with low frequency but large amplitude~\cite{Zhang2005}. Due to the small magnitude of $\beta$, typical spin-dependent mean-field energy is of the order of 10 Hz, and as a result, visible heteronuclear spin oscillations can only occur near $\Delta E = 0$. As shown in Fig.~\ref{fig:principle}, $\Delta E$ depends on the magnetic field $B$ in a non-monotonic manner, and in particular vanishes at a field $B_0=1.69$ G, where one expects resonant spin dynamics. This coincidence is a result of the slightly different Land\'e g-factors for $^{23}$Na and $^{87}$Rb, including contributions from both the linear and the quadratic Zeeman energies~\cite{Supp}. Near $B_0$, homonuclear spin dynamics is greatly suppressed due to large quadratic Zeeman shifts and the heteronuclear spin dynamics for otherwise allowed spin changing processes are also suppressed due to large detuning. For example, the magnetic energy difference for the spin changing process $\ket{0,1} \leftrightarrow \ket{1,0}$ is larger than $1000$ Hz and will be substantially suppressed. Thus, working near $B_0$, we can single out the process in Eq.~(\ref{eqn_spin_exchange}) and obtain clear signatures of heteronuclear spin dynamics.

The considerations above offer only a qualitative picture of the inter-species spin dynamics. Experimentally, we use a bulk sample consisting of an essentially pure $^{23}$Na BEC and a thermal gas of $^{87}$Rb to increase the overlap of the two clouds. This many-body system is distinctively different from the conventional two-level system since spin- and density-dependent mean-field interactions enter nonlinearly into the equations of motion and furthermore, vary in the course of spin dynamics~\cite{Supp}. One of the important consequences is the appearance of two magnetically tuned resonances as we shall discuss momentarily.

We produce the ultracold mixture of $^{23}$Na and $^{87}$Rb atoms in a crossed ODT with both atoms initially prepared in the spin state $\ket{-1}$~\cite{Wang2013,Xiong2013}. To initiate the spin oscillations, we apply a radio-frequency (rf) Rabi pulse to simultaneously prepare a coherent superposition state with most population in $\ket{-1}$ and $\ket{0}$ for both Rb and Na, while populations in the $\ket{+1}$ states are typically less than 10$\%$. To monitor the spin dynamics, we detect the fractional spin population $\rho^{i}_{m} = N^{i}_{m}/N^{i}$ for each species from the absorption images after various holding time. Here $N^{i}_{m}$ is the atom number of species $i$ in spin state $\ket{m}$. $N^{i} = N^{i}_{-1}+N^{i}_{0}+N^{i}_{+1}$ is the total number of atoms of species $i$, with $i={\rm Na,Rb}$.

\begin{figure}
\includegraphics[width=0.9 \linewidth]{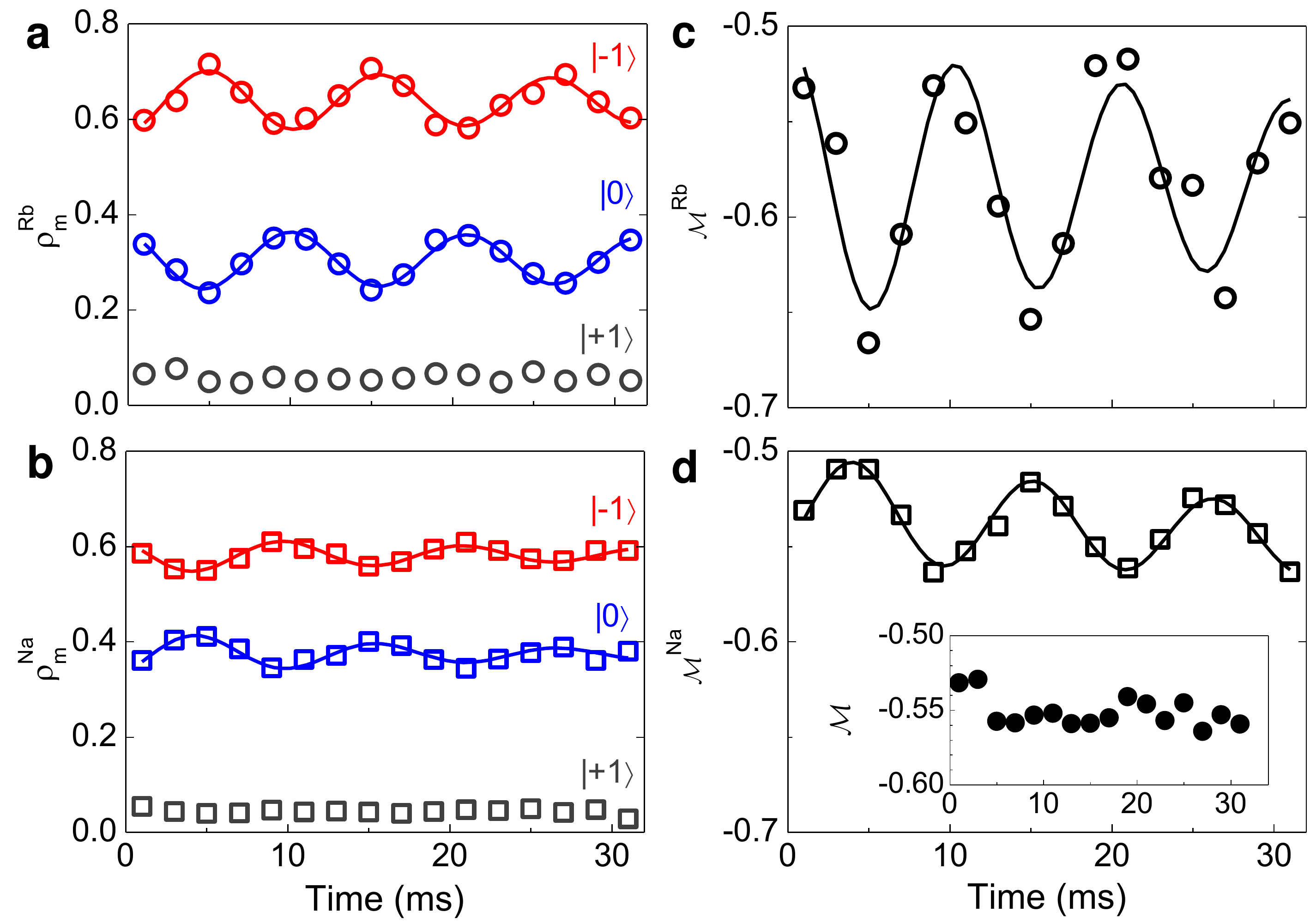}
\caption{\label{fig:oscillation} (color online) Coherent heteronuclear spin dynamics at $B=1.9$ G. {\bf a}, {\bf b}, evolution of the fractional spin populations of spin-1 Rb (circle) and Na (square). Red, blue and gray colors label the three Zeeman states $\ket{-1}$, $\ket{0}$, and $\ket{+1}$, respectively. {\bf c}, {\bf d},  magnetization oscillations of Rb and Na show $\pi$-phase shift. The differences in the oscillation amplitudes of Rb and Na are due to their number imbalance and the conservation of the total magnetization, which, as shown in the inset, remains approximately constant in the course of spin dynamics. All solid lines are sinusoidal fitting to experimental data (see Methods).}
\end{figure}

\begin{figure*}
\includegraphics[width=0.9\linewidth]{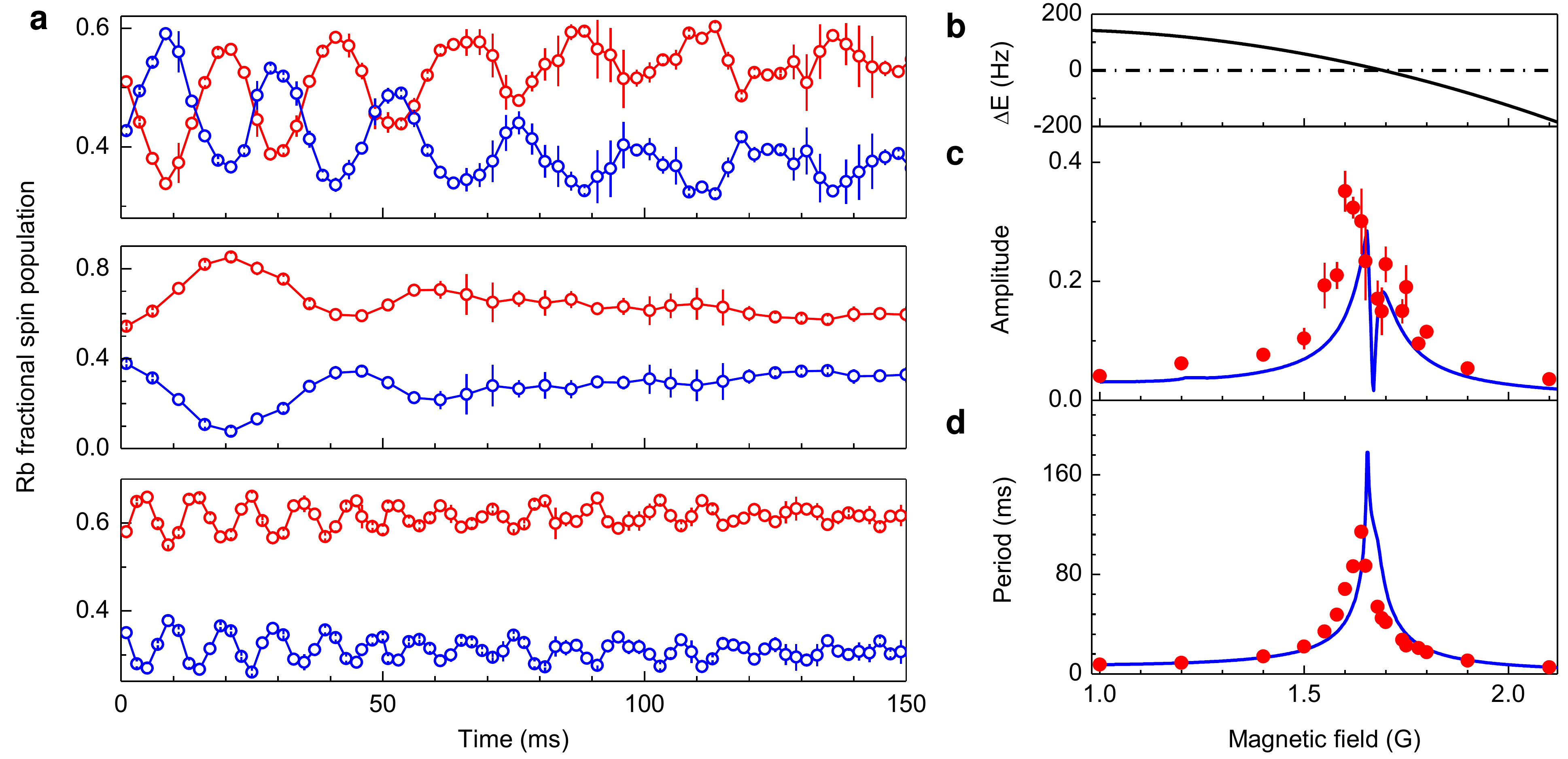}
\caption{\label{fig:resonance} (color online) Dependences of heteronuclear spin dynamics on external magnetic field $B$. {\bf a}, spin oscillations for Rb atom at $B=1.5$ G (top), $1.7$ G (middle) and $1.9$ G (bottom). The fractional spin populations in $\ket{-1}$ (red) and $\ket{0}$ (blue) are monitored over long time. Away from resonance at $B=1.5$ G and $1.9$ G, many oscillations are observed with short period and  small amplitude. Close to resonance at $B=1.7$ G, however, oscillations are slow but with larger amplitude.  The solid curves are for eye guiding and error bars are from statistics of several shots. The population in $\ket{+1}$ state is in general less than $10\%$ and furthermore remains constant in the course of spin dynamics, so it is not shown here. {\bf b}, magnetic energy $\Delta E$ as a function of $B$ with zero crossing at $B_0=1.69$ G.  {\bf c}, {\bf d},  spin oscillation amplitudes {\bf c} and periods {\bf d} extracted from the experimental data and two resonances can be clearly identified in the amplitude. Solid blue lines are calculations based on many-body kinetic equations using experimental atomic conditions without fitting parameters. Error bars for both the amplitude and the period are from fitting of the oscillations and represent one standard deviation(see Methods). Mechanisms for the observed damping in the oscillations will be investigated in future works.}
\end{figure*}

Fig.~\ref{fig:oscillation}a and ~\ref{fig:oscillation}b show typical time evolution of $\rho^{\rm Rb}_{m}$ and $\rho^{\rm Na}_{m}$ at $B=1.9$ G, respectively. The population in states $\ket{-1}$ and $\ket{0}$ oscillate periodically, while those in state $\ket{1}$ stays nearly constant. It is important to note the following features: (1) States $\ket{-1}$ and $\ket{0}$ of each individual species oscillate with $\pi$-phase shift due to number conservation; (2) The synchronized oscillation between the two species reflects the coherent spin dynamics driven by heteronuclear spin exchange interaction. This is even more clearly exhibited in the magnetization dynamics of the two species. The fractional magnetization for each species is $\mathcal{M}^{i} = (N_{+1}^{i}-N_{-1}^{i})/N^{i}$. The total magnetization of the system is defined as $\mathcal{M}= (N^{Na}_{+1}-N^{Na}_{-1}+N^{Rb}_{+1}-N^{Rb}_{-1})/N$, where $N=\sum_i N^i$ is the total number of atoms. As shown in Fig.~\ref{fig:oscillation}c, ~\ref{fig:oscillation}d and the inset, $\mathcal{M}^{\rm Na}$ and $\mathcal{M}^{\rm Rb}$ are not conserved, but total $\mathcal{M}$ is conserved within a few percent. The coherent oscillations of $\mathcal{M}^{\rm Na}$ and $\mathcal{M}^{\rm Rb}$ with a $\pi$-phase difference is a clear signature of the coherent heteronuclear spin exchanging process. The clean oscillations between the $\ket{-1}$ and $\ket{0}$ states also indicates that homonuclear and the other heteronuclear spin changing processes are greatly suppressed.

Similar measurements of spin dynamics are performed for a range of magnetic fields; three examples for Rb are plotted in Fig.~\ref{fig:resonance}a. Away from $B_0$, fast oscillation with small amplitude can be observed, while very close to $B_0$, {\em e.g.} at $B=1.7$ G (middle), the oscillation is slow but with large amplitude. One further feature is worth noticing. Compare oscillations at $B=1.5$ G (top) and 1.9 G (bottom), we note that the initial slopes of population change for the same spin states have opposite signs on different sides of $B_0$. This is a direct reflection of the sign change in $\Delta E$, as depicted in Fig.~\ref{fig:principle}. Similar behavior is observed for Na. These correlated spin oscillations for two species are well reproduced in our numerical simulations and furthermore are consistent with the ferromagnetic exchange interaction $\beta<0$.

We extract the oscillation amplitudes and periods for different magnetic fields and summarize the results in Fig.~\ref{fig:resonance}c and ~\ref{fig:resonance}d. Near $B_0$, the system is in the interaction dominated regime where an asymmetric double peak appears in the oscillation amplitude with a non-zero dip in between. This can be understood by noting that resonances appear when the absolute values of $\Delta E$ and spin-dependent interaction are comparable, which can occur on either side of $B_0$, analogous to the single species case where $\Delta E$ is tuned by quadratic Zeeman shift~\cite{Zhao2014}. However, the exact resonance positions depend also on homonuclear spin-dependent interactions and initial conditions. The double peak behavior is, however, not readily distinguishable in the period where only one peak is observed~\cite{Supp}.

To understand the observed spin dynamics quantitatively, we model the Na condensate with the time dependent Gross-Pitaevskii equation~\cite{Law1998,Xu2012} and the thermal Rb cloud with the kinetic equation for the Wigner distribution function~\cite{Endo2008,Natu2010,Pechkis2013}. The dynamics of the two species are coupled through the interaction in Eq.(\ref{eq:interaction}). Within the random phase and single mode approximations~\cite{Supp}, our simulation agrees well with the measurements, as shown in Fig.~\ref{fig:resonance}c and ~\ref{fig:resonance}d. The simulated oscillation frequency shows only a small kink near $B_0$, in accordance with the experiment.

\begin{figure}
\includegraphics[width=0.9\linewidth]{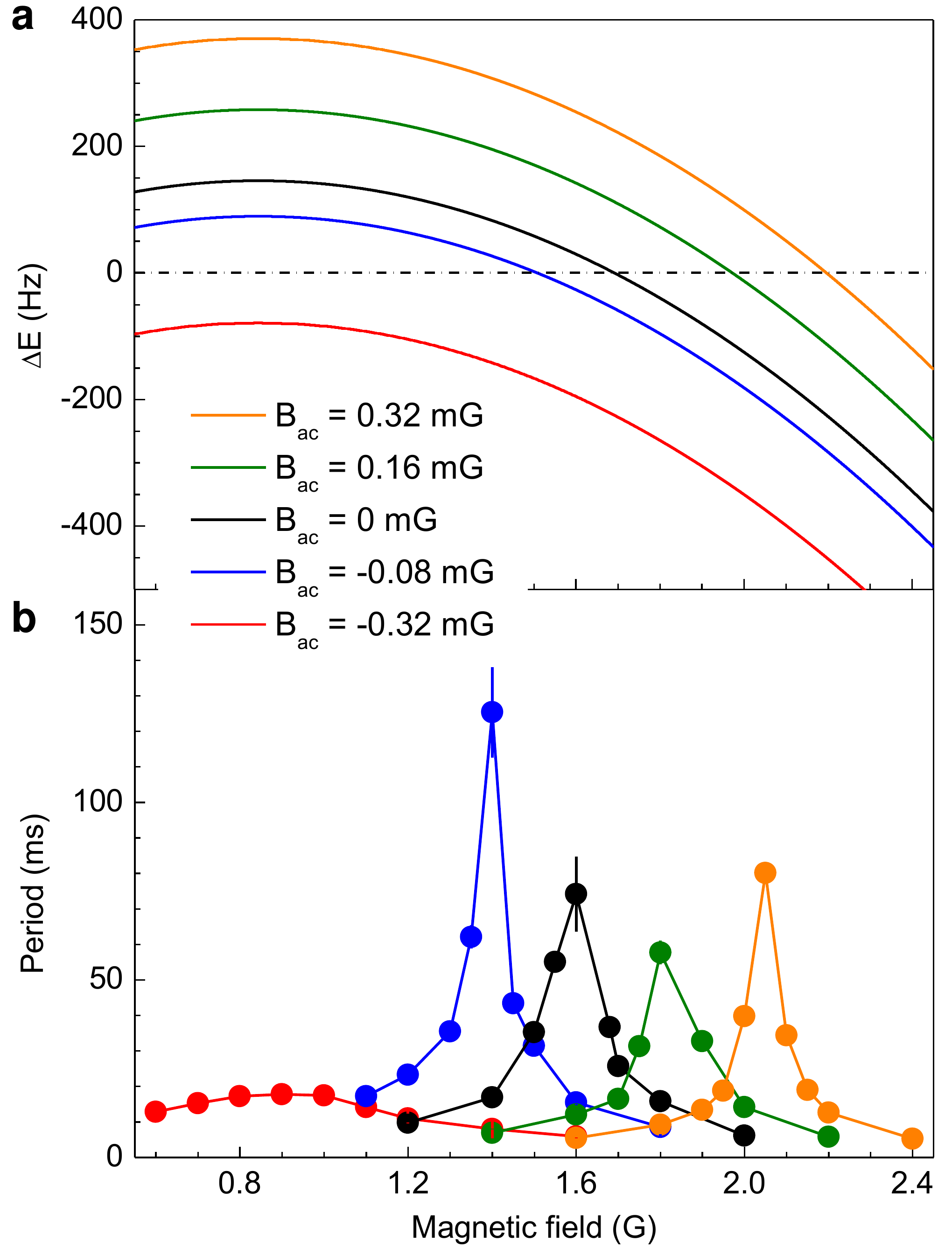}
\caption{\label{fig:tuning} (color online) Optical control of coherent heteronuclear spin dynamics with vector light shift. {\bf a}, modified dependence of $\Delta E$ for Rb on $B$ when light induced effective magnetic field $B_{\rm ac}$ is taken into account. Zero crossing of $\Delta E$ moves towards high magnetic fields $B$ as $B_{\rm ac}$ increases; while for sufficiently negative $B_{\rm ac}$, zero crossing vanishes. {\bf b}, resonance positions as observed in the period vary with changing $B_{\rm ac}$, and follow the locus of the minimum of $|\Delta E|$. Here the external magnetic field $B$ is applied horizontally with a $31^{\circ}$ angle to the trap beams. Solid curves are for eye guiding and error bars are from fitting of the oscillations. $B_{\rm ac}$ is calculated based on the measured light intensity $I$ and $\wp$.}
\end{figure}

A unique feature of the heteronuclear spin dynamics is its dependence on the vector light shift, which is spin- and species-dependent~\cite{Supp}. In the following, we tune the ellipticity of the ODT beams to further control the coherent spin dynamics. In the case of large detuning $\Delta$, exceeding the excited state fine structure splitting $\Delta_{\rm FS}$, the spin-dependent vector light shift is~\cite{Grimm2000}
\begin{equation}
U_m(\vec{r}) \propto \frac{\wp m}{\omega^{3}_{0}}\frac{\Delta_{\rm FS}}{\Delta^2}I(\vec{r}),
\label{eq:potential}
\end{equation}
where $\omega_{0}$ is the energy splitting between the ground state and the center of the $D$-lines, and $I(\vec{r})$ is the light intensity. The factor $\wp$ characterizes the amount of circular polarization with $\wp = 0$ for linear and $\wp = \pm 1$ for pure $\sigma^{\pm}$ circular polarizations, respectively. $U_m$ can be treated as a``fictitious magnetic field'' in the light propagation direction~\cite{Cohen1972}. Its projection, $B_{\rm ac}$, along the quantization axis alters the effective magnetic field seen by the atoms. Due to the larger $\Delta$, $\omega_0$ and smaller $\Delta_{\rm FS}$ for $^{23}$Na, $B_{\rm ac}$ for $^{23}$Na is more than hundred times smaller than that for $^{87}$Rb. For our final ODT, $B_{\rm ac} \approx$ 1.6 mG for Rb and 14 $\mu$G for Na if $\wp = 1$. So effectively speaking, by tuning $\wp$, we can control the linear Zeeman energy for Rb and Na independently~\cite{Supp}. The measurements shown in Fig.~\ref{fig:oscillation} and Fig.~\ref{fig:resonance} are performed with $B_{\rm ac}$ essentially zero.

Although small, $B_{\rm ac}$ has a dramatic influence on the heteronuclear spin dynamics. For simplicity, the much smaller $B_{\rm ac}$ for Na is ignored from now on. On the other hand, the much larger $B_{\rm ac}$ for $^{87}$Rb can shift $\Delta E$ significantly, as illustrated in Fig.~\ref{fig:tuning}a. For $B_{\rm ac}<0$, the zero crossing point is shifted to smaller external magnetic fields. Eventually, for $B_{\rm ac} < -0.2$ mG the entire $\Delta E$ curve is shifted to below zero and the zero crossing disappears. In such cases, the spin dynamics will be essentially driven by Zeeman energies with a peak at the field of minimum $|\Delta E|$. When $B_{\rm ac}>0$, the zero crossing point and thus the resonance position always shifts to higher magnetic field.

Experimentally, $\wp$, and hence $B_{\rm ac}$, can be tuned by applying the external magnetic field in the horizontal plane and inserting a $\lambda/4$ waveplate into one of the ODT beams. Here $\wp$ = sin(2$\theta$), where $\theta$ is the angle between the waveplate's axis and the input linear polarization of the light. For the typical range of $\theta$ varied in our experiment without causing significant heating, $B_{\rm ac}$ ranges from $-0.32$ to $0.32$ mG. As shown in our measurement in Fig.~\ref{fig:tuning}b, a rather small $B_{\rm ac}$ can cause a significant change of the resonance position. For example, at $B_{\rm ac}= 0.32$ mG, the resonance is shifted upwards by about $0.4$ G. On the other hand, for negative $B_{\rm ac}$ such that the zero crossing disappears, the lineshape of the oscillation becomes much broader, as for example when $B_{\rm ac}=-0.32$ mG, where the oscillation is always far off resonance.

In conclusion, we have observed interaction driven coherent spin changing dynamics between two different spin-$1$ Bose gases. Both the oscillation period and amplitude can be tuned over a large range with either external magnetic fields or, quite unique to our system, the light polarizations of the ODT. This latter capability is especially promising because it allows sensitive and versatile control of the spin dynamics, as demonstrated in our experiment. Our system may also serve as an ideal platform for simulating complicated spin dynamics in solid state physics, such as coupled electronic and nuclear spin systems.

\section{Acknowledgments}
We thank Bo Gao for the scattering lengths calculation, Li You, Qi Zhou, Hui Zhai and Ana Maria Rey for stimulating discussions, and C.K.Law for reading the manuscript. This work is supported by Hong Kong Research Grants Council (General Research Fund Projects CUHK 403813 and CUHK 14305214). Z.F.X. is supported by AFOSR and ARO. S.Z.Z. is supported by Hong Kong Research Grants Council (General Research Fund, HKU 709313P and Collaborative research fund, HKUST3/CRF/13G).

\end{document}

\title{Supplementary Material: Coherent heteronuclear spinor dynamics in an ultracold spin-1 mixture}

\author{Xiaoke Li}
\affiliation{Department of Physics, The Chinese University of Hong Kong, Hong Kong, China}
\author{Bing Zhu}
\affiliation{Department of Physics, The Chinese University of Hong Kong, Hong Kong, China}
\author{Xiaodong He}
\affiliation{Department of Physics, The Chinese University of Hong Kong, Hong Kong, China}
\author{Fudong Wang}
\affiliation{Department of Physics, The Chinese University of Hong Kong, Hong Kong, China}
\author{Mingyang Guo}
\affiliation{Department of Physics, The Chinese University of Hong Kong, Hong Kong, China}
\author{Zhi-Fang Xu}
\affiliation{Department of Physics and Astronomy, University of Pittsburgh, Pittsburgh, Pennsylvania 15260, USA}
\author{Shizhong Zhang}
\affiliation{Department of Physics and Centre of Theoretical and Computational Physics, The University of Hong Kong, Hong Kong, China}
\author{Dajun Wang}
\email{djwang@phy.cuhk.edu.hk}
\affiliation{Department of Physics, The Chinese University of Hong Kong, Hong Kong, China}

\date{\today}

\maketitle

\section{Experimental methods}
\subsection{Preparation of the spinor mixture}
We produce the ultracold mixture of an essentially pure $^{23}$Na BEC with 1.0$\times$10$^5$ atoms and a $^{87}$Rb thermal gas with about $6.3\times$10$^4$ atoms in a crossed ODT, following the procedure described in~\cite{Xiong2013}. Thermal $^{87}$Rb is used here to increase the overlap between two species as double BEC is immiscible~\cite{Xiong2013}. The number imbalance is chosen such that the $^{87}$Rb cloud is $\sim$100 nK above its BEC transition temperature. The final stage of the evaporation is performed in the presence of a $2$ G magnetic field to make sure that all atoms are polarized in the $\ket{-1}$ spin state. The final trap frequencies for Rb and Na are $2\pi\times (110, 215, 190)$ Hz and $2\pi\times(98, 190, 168)$ Hz, respectively. The average density is 5.9$\times$10$^{13}$ cm$^{-3}$ (6.5$\times$10$^{12}$ cm$^{-3}$) for Na (Rb). Along the vertical direction, the Thomas-Fermi radius of the Na BEC is $7.1\mu$m and the size of the thermal Rb cloud is $5.1\mu$m. There is a differential gravitational sag of about $2.4\mu$m due to the trap frequency difference in the vertical $y$-direction.

The magnetic field is then tuned to values near $B_0$, which are calibrated to within $\pm 1$ mG with microwave Rabi spectroscopy performed on Rb. A rf pulse with Rabi frequency much larger than the quadratic Zeeman energies is applied afterward for spin states initialization. Three-level Rabi oscillations are observed for both species simultaneously and we choose a pulse duration of $10\mu$s, which corresponds to a pulse area of $\pi/3$, in order to put most of the populations in the $\ket{-1}$ and $\ket{0}$ states.

\subsection{Spin population measurement}
After holding for various durations, we switch off the ODT and detect the spin population of each species with spin-selective Stern-Gelarch time-of-flight absorption imaging, by applying a magnetic field gradient during the expansion. For each experimental run, we take two images, one for each species. To obtain accurate atom numbers, our image system is calibrated for both Na and Rb~\cite{Reinaudi2007,Kwon2012}.

\subsection{Fitting of the spin oscillations}
Spin oscillations have the same period for Na and Rb. But due to the smaller number of Rb and the the fact that total $\mathcal{M}$ is conserved, the fractional populations of Rb show larger oscillation amplitude than Na and offers better signal to noise ratio. We thus fit the Rb oscillation only with a damped sinusoidal function of amplitude $A_i$ for various spin components and period $T$
\begin{equation}
\rho_{i}(t)=C_i + A_ie^{-t/\tau}\sin(2\pi t/T+\varphi),
\label{eq:fit}
\end{equation}
where $C_i$ is an offset determined by the initial state. $\tau$ is the damping constant and $\varphi$ is the initial phase of the oscillation. Close to $B_0$, oscillations are typically strongly damped, in which case, only the first few periods are used in the fitting.

\section{Zeeman shift and light induced magnetic field}
\subsection{Zeeman shift}
The Zeeman shifts for both $^{87}$Rb and $^{23}$Na can be expressed as
\begin{equation}
\begin{array}{c}
{\hat H_Z} =  - p{\hat F_z} + q\hat F_z^2,
\end{array}
\end{equation}
where $p$ and $q$ are the linear and quadratic Zeeman shifts, respectively. Both of them can be obtained from the power series expansion of the Breit-Rabi formula. The fine structure and nuclear Land\'e g-factors are $g_{J} = 2.00233113\left( {2.0022960} \right)$ and $g_{I} =  - 0.0009951414\left( { - 0.0008046108} \right)$ for Rb(Na)~\cite{Arimondo1977}. The different values of parameters ($g_{J}$ and $g_{I}$) for Na and Rb renders different linear as well as quadratic Zeeman shifts ($p$ and $q$), and we have to keep both terms in the theoretical treatment.

With these g-factors, the Zeeman energy difference $\Delta E\left( B \right)$ (see main text) is calculated. For the spin exchanging process $\ket{0,-1}\leftrightarrow\ket{-1,0}$ which we are focusing on in this work, $\Delta E = 0$ at 1.69G.

In the presence of light induced effective magnetic field $B_{\rm{ac}}$, the $\Delta E = 0$ point will be shifted to new values determined by replacing $B$ with $B+B_{\rm{ac}}$ for Rb. As discussed in the main text, the much smaller light induced $B$ field on Na can be ignored. For Rb, $p_{\rm{Rb}}/B \approx 702.369 \left( \rm{kHz} \cdot \rm{G}^{- 1} \right)$. As $B_{\rm{ac}}$ is typically small in our system, its contribution on $q$ can also be ignored.

\subsection{Calculation of light induced effective B field}

As the frequency of our ODT laser is red detuned with respect to the Rb and Na D-lines with detunings $\Delta$ much larger than the fine structure splitting $\Delta_{\rm FS}$, the optical trap potential can be approximated as\cite{Grimm2000}
\begin{equation}
U_{\rm dipole}({\bf r}) = U_{\rm scaler}({\bf r}) + U_{\rm m}({\bf r})
=  \frac{3 \pi c^2}{2 \omega_0^3} \,
\frac{\Gamma}{\Delta} \,
\left( \, 1 \, + \, \frac{1}{3} \, {\wp} g_f m \,
\frac{\Delta_{\rm FS}}{\Delta} \, \right)
\, I({\bf r}) \, .
\label{ODT1}
\end{equation}
Here c is the speed of light and $g_f$ is the hyperfine Land\'e g-factor. $\Gamma$ and $\omega_0$ are the natural linewidth and transition frequency of the $D$-lines, respectively. For the $f$ = 1 hyperfine state, we have m = 0 and $\pm 1$. $U_{\rm scaler}$ is from the scalar polarizability which causes an overall shift to all m-states. $U_{\rm m}$ is from the vector polarizability and is m-dependent. It is only present when the circular polarization component $\wp$ is not zero. Effectively, the vector potential can be treated as a magnetic field~\cite{Cohen1972}.

Our ODT is formed by two linear and orthogonally polarized beams propagating in the horizontal plane and crossing each other at an angle of 62$\degree$. To generate the effective magnetic field, we insert a $\lambda/4$ waveplate into one of the ODT beams. The crystal axis of the waveplate is first aligned with the light polarization. Then by rotating the waveplate, we can control $\wp = {\rm sin}(2\theta)$, with $\theta$ the angle of rotation.

Experimentally, $\theta$ has been limited to $\pm{\rm 6}\degree$ as excessive heating occurs at larger $\theta$, probably due to interference between the two ODT beams. After obtaining the light intensity from the measured laser power and beam waist, we calculate the effective $B$ field as
\begin{equation}
B_{\rm ac} = {\rm cos}{\phi}\times U_{\rm m}/g_f m \mu_{\rm B},
\label{Bac}
\end{equation}
with $\phi$ the angle between the laser beam and the quantization axis defined by the externally applied magnetic field. Here $\mu_{\rm B}$ is the Bohr magneton. When the quantization axis is along the vertical direction, $\phi = 90\degree$ and thus $B_{\rm ac}$ is essentially zero. This is the case for the measurements presented in Fig. 2 and Fig. 3 of the main text. When the quantization axis is in the horizontal direction, $\phi = 31\degree$ and thus we can have a non-zero $B_{\rm ac}$ for finite $\wp$.

\section{Some parameters used in the calculation}
\label{parameters}
In this section, we give explicit expression for some parameters that enter the theoretical simulations.
\subsection{scattering lengths}
For the homonuclear cases, the total spin can only be $F$ = 0 and 2. As listed in Table~\ref{table1}, $a_0$ and $a_2$ are taken from~\cite{Klaussen2001} for Rb and ~\cite{Crubellier1999} for Na. The spin-dependent interaction is determined by the scattering length difference $\Delta a = a_2 - a_0$. We notice that uncertainties remain in $\Delta a$ as reported by several groups~\cite{Chang2005,Widera2006,Stenger1998,Black2007}. The heteronucelar scattering lengths in Table~\ref{table1} are calculated based on the electronic singlet and triplet scattering lengths, and the molecular potentials in~\cite{Wang2013,Gao2014}. The uncertainties of these values are on the one percent level.
\begin{table}[h]
\begin{tabular}{c | c |c|c}
\hline
\hline
	& Rb-Rb & Na-Na & Rb-Na  \\
\hline
$F=0$& $a_{\rm Rb}^{(0)}=101.8$ & $a_{\rm Na}^{(0)}=47.36$ & $a_{\rm Rb-Na}^{(0)}=83.81$\\
\hline
$F=1$&  /  & / &$a_{\rm Rb-Na}^{(1)}=81.37$\\
\hline
$F=2$& $a_{\rm Rb}^{(2)}=100.4$ & $a_{\rm Na}^{(2)}=52.98$ & $a_{\rm Rb-Na}^{(2)}=76.38$\\
\hline
\end{tabular}
\caption{Homo- and heteronuclear total spin scattering lengths used in our simulation. All the scattering lengths are given in $a_B$, the Bohr radius.}
\label{table1}
\end{table}

The uncertainty in these scattering lengths is one of the main causes of the small discrepancy between theory and experiment in Fig. 3c and Fig. 3d of the maintext.

\subsection{Interaction parameters}
For the homonuclear case, the two-body interaction can be expressed as
\begin{equation}
{V_{12}}\left( {\bf{r_1} -\bf{r_2} } \right) = \left( {c_{\rm Na,Rb}^{(0)}  + c_{\rm Na,Rb}^{(2)} {{\bf{f}}_1} \cdot {{\bf{f}}_2} } \right)\delta \left( {\bf{r_1} -\bf{r_2}} \right).
\label{homo}
\end{equation}
The interaction parameters $c_{\rm Na,Rb}^{(0,2)}$ can be written as
\begin{align}
c_{\rm Na,Rb}^{(0)} &=\frac{4\pi\hbar^2}{m_{\rm Na,Rb}}\frac{a_{\rm Na,Rb}^{(0)}+2a_{\rm Na,Rb}^{(2)}}{3},\\
c_{\rm Na,Rb}^{(2)} &=\frac{4\pi\hbar^2}{m_{\rm Na,Rb}}\frac{a_{\rm Na,Rb}^{(2)}-a_{\rm Na,Rb}^{(0)}}{3}.
\end{align}
The parameters $\a,\b$ and $\g$ can be written as
\begin{align}
\a  &= \frac{2\pi\hbar^2}{\mu}\frac{a_{\rm Rb-Na}^{(1)}+a_{\rm Rb-Na}^{(2)}}{2},\\
\b  &= \frac{2\pi\hbar^2}{\mu}\frac{-a_{\rm Rb-Na}^{(1)}+a_{\rm Rb-Na}^{(2)}}{2},\\
\g  &= \frac{2\pi\hbar^2}{\mu}\frac{2a_{\rm Rb-Na}^{(0)}-3a_{\rm Rb-Na}^{(1)}+a_{\rm Rb-Na}^{(2)}}{2},
\end{align}
where $\mu=m_{\rm Na}m_{\rm Rb}/(m_{\rm Na}+m_{\rm Rb})$ is the reduced mass.

\section{Theoretical calculations}
In the main text, a simple picture based on two-level system is introduced. Such a picture is helpful for establishing an intuitive understanding of the spin dynamics, but cannot catch all the many-body effects. For a complete description, we have developed a mean-field model. Here both the two-body and the many-body theories are presented.

\subsection{Two-body theory}
Let's focus on the spin changing process $\ket{0,-1} \leftrightarrow \ket{-1,0}$, and ignore influences of all other processes. As $M = m_1 + m_2 = -1$ is conserved, the above spin changing process involves only total spin $F = 1$ and $2$ channels. The spin-dependent interaction Hamiltonian can then be expressed as
\begin{equation}
H_{\rm I}=\sum_{F}g_{F}\ket{F,M}\bra{F,M}=g_2\ket{2,-1}\bra{2,-1}+g_1\ket{1,-1}\bra{1,-1}.
\end{equation}
Evaluate the Clebsch-Gordan coefficient between $\ket{m_1,m_2}$ and $\ket{F,M}$, we have
\begin{align}
&\ket{2,-1}=\frac{1}{\sqrt{2}}(\ket{0,-1}+\ket{-1,0}) \\
&\ket{1,-1}=\frac{1}{\sqrt{2}}(\ket{0,-1}-\ket{-1,0}).
\end{align}
Thus in the basis of $\ket{0,-1}$ and $\ket{-1,0}$, $H_{\rm I}$ can be expressed as
\begin{align}
H_{\rm I}=\frac{1}{2}\left[\begin{array}{cc}
g_2+g_1 & g_2-g_1\\
g_2-g_1 & g_2+g_1
\end{array}\right],
\end{align}
where $(g_2-g_1)/2$ is exactly the spin-dependent $\beta$ term. The Zeeman energy can be written as
\begin{align}
H_{\rm Z}=\left[\begin{array}{cc}
E_1(B) & 0\\
0 & E_2(B)
\end{array}\right],
\end{align}
and the total Hamiltonian is
\begin{align}
H = H_{\rm I} + H_{\rm Z}=\left[\begin{array}{cc}
(g_2+g_1)/2+E_1(B) & (g_2-g_1)/2\\
(g_2-g_1)/2 & (g_2+g_1)/2+E_2(B)
\end{array}\right].
\end{align}
The generalized Rabi frequency of the two-particle system is given by
\begin{equation}
\Omega = \sqrt{(E_1(B)-E_2(B))^2+(g_2-g_1)^2}/\hbar=\sqrt{\Delta E(B)^2+(g_2-g_1)^2}/\hbar.
\end{equation}
At $B_0 = 1.69$ G, where $\Delta E(B) = 0$, this model predicts resonant oscillation with the longest period and largest amplitude. This result disagrees with our observation apparently.

On either side of $B_0$, we should have detuned oscillations with shorter periods and smaller amplitudes. When $\Delta B$ is much larger than $g_2-g_1$, the amplitude becomes too small to be observed. Our measurements in this region agree with this prediction qualitatively. The oscillation period can be well approximated as $1/\Omega$.

\subsection{Many-body theory}
\label{mbt}
In this part, we outline the basic kinetic approach that we employ to interpret the experimental data. For $^{87}$Rb  thermal cloud, the evolution of spin polarization as well as spatial distribution can be described, within semiclassical approximation, with the so-called Wigner function
\begin{equation}
g_{ij}({\bf R}, {\bf p};t)\equiv \int d^3{\bf r}e^{-i{\bf p}\cdot{\bf r}}\left\langle\psi^\dagger_{j}({\bf R}+\frac{\bf r}{2},t)\psi_{i}({\bf R}-\frac{\bf r}{2},t)\right\rangle,
\end{equation}
where $\psi_{i}({\bf r})$ is the Heisenberg annihilation operator for Rb atom with spin $\a$ and at position ${\bf r}$. In our calculation, all three spin components of the $F=1$ manifold of the $^{87}$Rb are taken into account and as a result, $g_{ij}({\bf R}, {\bf p};t)$ is a $3 \times 3$ matrix. In the following, we shall label the Zeeman levels by $i,j=-1,0,1$ for both Na and Rb.  The equation that governs the time-dependences of the Wigner function $g_{ij}({\bf R}, {\bf p};t)$ can be written in the following general form~\cite{Endo2008}
\begin{equation}
\frac{\partial g({\bf R},{\bf p};t)}{\partial t}+\frac{{\bf p}}{m_{\rm Rb}}\cdot\nabla g({\bf R},{\bf p};t)-\frac{1}{2}\{\nabla_{\bf R} U_{\rm Rb}, \nabla_{\bf p} g({\bf R},{\bf p};t)\}-\frac{i}{\hbar}[g({\bf R},{\bf p};t),U_{\rm Rb}]=I_{\rm Rb}.
\label{rbeqn}
\end{equation}
where $I_{\rm Rb}$ is the collision integral that describes the effects of interactions that are not captured in the effective potential $U_{\rm Rb}({\bf R},{\bf p};t)$, which is in general a (matrix) function of ${\bf R},{\bf p}$ and $t$. In our calculation, $U_{\rm Rb}({\bf R},{\bf p};t)$ is obtained within the random phase approximation (RPA), which we generalize to the case of Bose-Bose mixtures. The anti-commutator $\{A,B\}$ and the commutator $[A,B]$ refer to quantities in spin space.

For the Na condensate, we shall make use of the mean-field Gross-Pitaevskii approximation. Let the field operator for the Na condensate to be $\phi_i({\bf r},t)$, then we can replace it by its expectation value $\phi_i({\bf r},t)\to \langle{\phi_i({\bf r},t)}\rangle$, which we will denote simply as $b_i({\bf r},t)$. To find the expression for $U_{\rm Rb}({\bf R},{\bf p};t)$ and the time dependences for $b_i({\bf r},t)$, we shall start with the full Hamiltonian, which can be written in three parts:
\begin{align}
H_{\rm Rb} &=\int d^3{\bf r} \left\{\psi_i^\dag\left[-\frac{\hbar^2\nabla^2}{2m_{\rm Rb}}\d_{ij}-p_{\rm Rb}F_z+q_{\rm Rb}F_z^2+V_{\rm Rb}\right]\psi_j+\psi_i^\dag\psi_j^\dag\left[\frac{c_{\rm Rb}^{(0)}}{2}\d_{il}\d_{jk}+\frac{c_{\rm Rb}^{(2)}}{2}{\bf F}_{il}\cdot{\bf F}_{jk}\right]\psi_k\psi_l\right\},\\
H_{\rm Na} &=\int d^3{\bf r} \left\{\phi_i^\dag\left[-\frac{\hbar^2\nabla^2}{2m_{\rm Na}}\d_{ij}-p_{\rm Na}F_z+q_{\rm Na}F_z^2+V_{\rm Na}\right]\phi_j+\phi_i^\dag\phi_j^\dag\left[\frac{c_{\rm Na}^{(0)}}{2}\d_{il}\d_{jk}+\frac{c_{\rm Na}^{(2)}}{2}{\bf F}_{il}\cdot{\bf F}_{jk}\right]\phi_k\phi_l\right\},\\
H_{\rm Rb-Na} &=\int d^3 {\bf r}\left\{\a\psi_i^\dag\psi_i \phi_j^\dag\phi_j+\b\psi^\dag_i{\bf F}_{il}\psi_l\cdot\phi^\dag_j{\bf F}_{jk}\phi_k+\g\frac{(-1)^{i-j}}{3}\psi^\dag_{i}\psi_{j}\phi_{-i}^\dag\phi_{-j}\right\}.
\end{align}
Here, $p_{\rm Rb,Na}$ and $q_{\rm Rb,Na}$ are the linear and quadratic Zeeman energy for Rb and Na atoms. $V_{\rm Rb,Na}$ are the confining harmonic trapping potential for the Rb and Na components, which we assume to be spin-independent. $c_{\rm Rb,Na}^{(0,2)}$ are the standard interaction parameters for the Rb and Na spinor gases~\cite{Kurn2013}. In the case of Bose-Bose mixture, inter-species interactions are described by three independent interaction parameters $\a,\b$ and $\g$~\cite{Xu2012}. Their explicit expressions in terms of scattering lengths are given in Sec.\ref{parameters}.

There are three different contributions to $U_{\rm Rb}({\bf R},{\bf p};t)$. The first comes from the single particle term
\begin{equation}
U^{(1)}_{{\rm Rb}}({\bf R})=V_{\rm Rb}({\bf R})-p_{\rm Rb}F_z+q_{\rm Rb}F_z^2.
\end{equation}
The second term comes from the intra-species interactions. Here one uses the RPA approximation and obtain the following expression~\cite{Endo2008}
\begin{equation}
U^{(2)}_{{\rm Rb}}({\bf R};t)=c_0^{\rm Rb}({\rm Tr}n_{\rm Rb}+n_{\rm Rb})+c_2^{\rm Rb}{\rm Tr}({\bf F}n_{\rm Rb})\cdot{\bf F}+c_2^{\rm Rb}{\rm Tr}({\bf F}n_{\rm Rb}\cdot{\bf F}).
\end{equation}
where ${\bf F}=(F_x,F_y,F_z)$ are the spin-1 operators and $n_{\rm Rb}$ is defined to be
\begin{equation}
n_{{\rm Rb},ij}({\bf R},t)=\left\langle\psi^\dag_j({\bf R},t)\psi_i({\bf R},t)\right\rangle.
\end{equation}
The last contribution comes from the inter-species interactions and can be written as, within GP mean-field theory
\begin{equation}
U^{(3)}_{{\rm Rb}}({\bf R};t)=\a{\rm Tr}(n_{\rm Na})+\b{\rm Tr}({\bf F}n_{\rm Na})\cdot{\bf F}+\g\mathcal{U}_\phi,
\end{equation}
where we have defined
\begin{align}
n_{{\rm Na},ij}({\bf R},t)=b^*_j({\bf R},t)b_i({\bf R},t)
\end{align}
and
\begin{align}
\mathcal{U}_\phi=\frac{1}{3}\left[\begin{array}{ccc}
b^*_{-1}b_{-1} & -b^*_{-1}b_{0} & b^*_{-1}b_{1}\\
-b^*_{0}b_{-1} & b^*_{0}b_{0} & -b^*_{0}b_{1}\\
b^*_{1}b_{-1} & -b^*_{1}b_{0}& b^*_{1}b_{1}
\end{array}\right].
\end{align}

Similarly, we can write the equation of motion for the condensate order parameter $b^\dag=(b^*_{-1},b^*_{0},b^*_{1})$.
\begin{align}\label{naeqn}
i\hbar \frac{\partial}{\partial t}b &=\left[-\frac{\hbar^2\nabla^2}{2m_{\rm Na}}-p_{\rm Na}F_z+q_{\rm Na}F_z^2+V_{\rm Na}+c_0^{\rm Na}{\rm Tr}(n_{\rm Na})+c_2^{\rm Na}(b^\dag {\bf F}b)\cdot{\bf F}\right]b\\\nonumber
&+\a{\rm Tr}(n_{\rm Rb})b+\b{\rm Tr}(n_{\rm Rb}{\bf F})\cdot {\bf F}b +\g\mathcal{U}_\psi b.
\end{align}
Here $\mathcal{U}_{\psi}$ is defined to be
\begin{align}
\mathcal{U}_\psi=\frac{1}{3}\left[\begin{array}{ccc}
\langle\psi^\dag_{-1}\psi_{-1}\rangle & -\langle\psi^\dag_{-1}\psi_{0}\rangle & \langle\psi^\dag_{-1}\psi_{1}\rangle\\
-\langle\psi^\dag_{0}\psi_{-1}\rangle & \langle\psi^\dag_{0}\psi_{0}\rangle & -\langle\psi^\dag_{0}\psi_{1}\rangle\\
\langle\psi^\dag_{1}\psi_{-1}\rangle & -\langle\psi^\dag_{1}\psi_{0}\rangle& \langle\psi^\dag_{1}\psi_{1}\rangle
\end{array}\right].
\end{align}
Now, equations (\ref{rbeqn}) and (\ref{naeqn}) furnish a complete description of the Bose mixture system with one component being thermal and the other condensate. In general, they are not very easy to solve. As a result, we shall make use of the {\it single-mode approximation} in the following analysis. For the Rb thermal cloud, we write
\begin{equation}
n_{{\rm Rb},ij}({\bf R},{\bf p};t)=\mathcal{Z}^{-1}\exp\left[-\frac{p^2/2m_{\rm Rb}+V_{\rm Rb}({\bf R})}{k_{\rm B}T}\right]\sigma_{ij}(t).
\label{thermalansatz}
\end{equation}
Namely, the ${\bf R}$ and ${\bf p}$ dependences of $n_{{\rm Rb},ij}$ is given by thermal distribution, independent of time, while its spin dependence is given by $\s_{ij}(t)$. Here
\begin{equation}
\mathcal{Z}=\int d^3{\bf R}\int \frac{d^3{\bf p}}{(2\pi\hbar)^3}\exp\left[-\frac{p^2/2m_{\rm Rb}+V_{\rm Rb}({\bf R})}{k_{\rm B}T}\right]
\end{equation}
is the classical partition function. For Na condensate, we write
\begin{equation}
b_i({\bf R};t)=\sqrt{{\rm Tr}(n_{\rm Na}({\bf R}))}\zeta_i(t)\equiv \sqrt{n_c({\bf R})}\zeta_i(t),
\label{ansatzcondensate}
\end{equation}
where the total condensate density of Na, $n_c({\bf R})$, is independent of time. The spin part of the condensate wave function is given by $\zeta_i(t)$ which carries all the time-dependences. The matrix $\s$ and the spinor $\zeta_i$ satisfy the following conditions:
\begin{equation}
{\rm Tr}\s=1,~~~\sum_i\zeta_i^*\zeta_i=1.
\end{equation}
We shall use a Thomas-Fermi form for the condense density distribution in the harmonic trap (see Sec.\ref{parameters}),
\begin{equation}
n_c({\bf R})=n_c({\bf 0})\left(1-\frac{x^2}{R_x^2}-\frac{(y+y_0)^2}{R_y^2}-\frac{z^2}{R_z^2}\right),
\end{equation}
where $y_0=g/\omega_y^{\rm Rb}$ is due to gravitational field, and $R_x$, $R_y$ and $R_z$ are the Thomas-Fermi radius of the condensate cloud and $n_c({\bf 0})$ is the central density.

With these preparations, we can substitute eqn.(\ref{thermalansatz}) into eqn.(\ref{rbeqn}) and integrate over ${\bf R}$ and ${\bf p}$ on both sides. Let us neglect the collision integral $I_{\rm Rb}$, which contributes to the damping of the oscillations and will not be discussed in this work. The second and third terms in the left hand of eqn.(\ref{rbeqn}) vanishes upon integration over ${\bf R}$ and ${\bf p}$. As a result, we find
\begin{equation}
\frac{\partial \s}{\partial t}=\frac{i}{\hbar}[\s,M_{\rm TG}],
\label{mainRb}
\end{equation}
where we have defined
\begin{equation}
M_{\rm TG}=-p_{\rm Rb}F_z+q_{\rm Rb}F_z^2+c_2^{\rm Rb}\bar{n}{\rm Tr}({\bf F}\s)\cdot{\bf F}+c_2^{\rm Rb}\bar{n}{\rm Tr}({\bf F}\s\cdot{\bf F})+\b\bar{n}_{tc}\sqrt{\frac{N_{\rm Na}}{N_{\rm Rb}}}{\rm Tr}({\bf F}\tau)\cdot{\bf F}+\g\bar{n}_{tc}\sqrt{\frac{N_{\rm Na}}{N_{\rm Rb}}}\mathcal{U}_\zeta,
\end{equation}
where $N_{\rm Rb, Na}$ are the numbers of atoms for Rb and Na. $\bar{n}$ and $\bar{n}_{tc}$ are defined by
\begin{align}
\bar{n} &=\frac{1}{N_{\rm Rb}}\int d^3{\bf R}[{\rm Tr}(n_{\rm Rb})]^2,\\
\bar{n}_{tc} &=\frac{1}{\sqrt{N_{\rm Rb}N_{\rm Na}}}\int d^3{\bf R}{\rm Tr}(n_{\rm Rb})n_c({\bf R}).
\end{align}
We have also defined the matrix $\tau$ in terms of its matrix elements
\begin{equation}
\tau_{ij}=\zeta^*_j\zeta_i,~~{\rm Tr}\tau=1,
\end{equation}
and the matrix
\begin{align}
\mathcal{U}_\zeta=\frac{1}{3}\left[\begin{array}{ccc}
\zeta^*_{-1}\zeta_{-1} & -\zeta^*_{-1}\zeta_{0} & \zeta^*_{-1}\zeta_{1}\\
-\zeta^*_{0}\zeta_{-1} & \zeta^*_{0}\zeta_{0} & -\zeta^*_{0}\zeta_{1}\\
\zeta^*_{1}\zeta_{-1} & -\zeta^*_{1}\zeta_{0}& \zeta^*_{1}\zeta_{1}
\end{array}\right].
\end{align}
The equation for the condensate order parameter $b$ can be written in terms of $\tau$, in a similar form as for $\s$. By substitute eqn.(\ref{ansatzcondensate}) into eqn.(\ref{naeqn}) and integrate over ${\bf R}$, we find
\begin{align}
\frac{\partial }{\partial t} \tau= \frac{i}{\hbar}[\tau,M_{\rm BEC}],
\label{mainNa}
\end{align}
with
\begin{equation}
M_{\rm BEC}=-p_{\rm Na}F_z+q_{\rm Na}F_z^2+c_2^{\rm Na}\bar{n}_c{\rm Tr}({\bf F}\tau)\cdot{\bf F}+\b\bar{n}_{tc}\sqrt{\frac{N_{\rm Rb}}{N_{\rm Na}}}{\rm Tr}({\bf F}\s)\cdot{\bf F}+\g\bar{n}_{tc}\sqrt{\frac{N_{\rm Rb}}{N_{\rm Na}}}\mathcal{U}_\s,
\end{equation}
where $\mathcal{U}_{\s}$ is defined to be
\begin{align}
\mathcal{U}_\s=\frac{1}{3}\left[\begin{array}{ccc}
\s_{-1,-1} & -\s_{-1,0} & \s_{-1,1} \\
-\s_{0,-1} & \s_{0, 0} & -\s_{0,1}\\
\s_{1,-1} & -\s_{1,0} & \s_{1,1}
\end{array}\right]
\end{align}
and
\begin{align}
\bar{n}_c &=\frac{1}{N_{\rm Na}}\int d^3{\bf R}n_c({\bf R})^2.
\end{align}
Equations (\ref{mainRb}) and (\ref{mainNa}) describe how the spin evolves under the influences of the effective external ``fields", $M_{\rm TG}$ and $M_{\rm BEC}$.

\subsection{Numerical simulations}

\begin{figure}
\includegraphics[width=0.6 \linewidth]{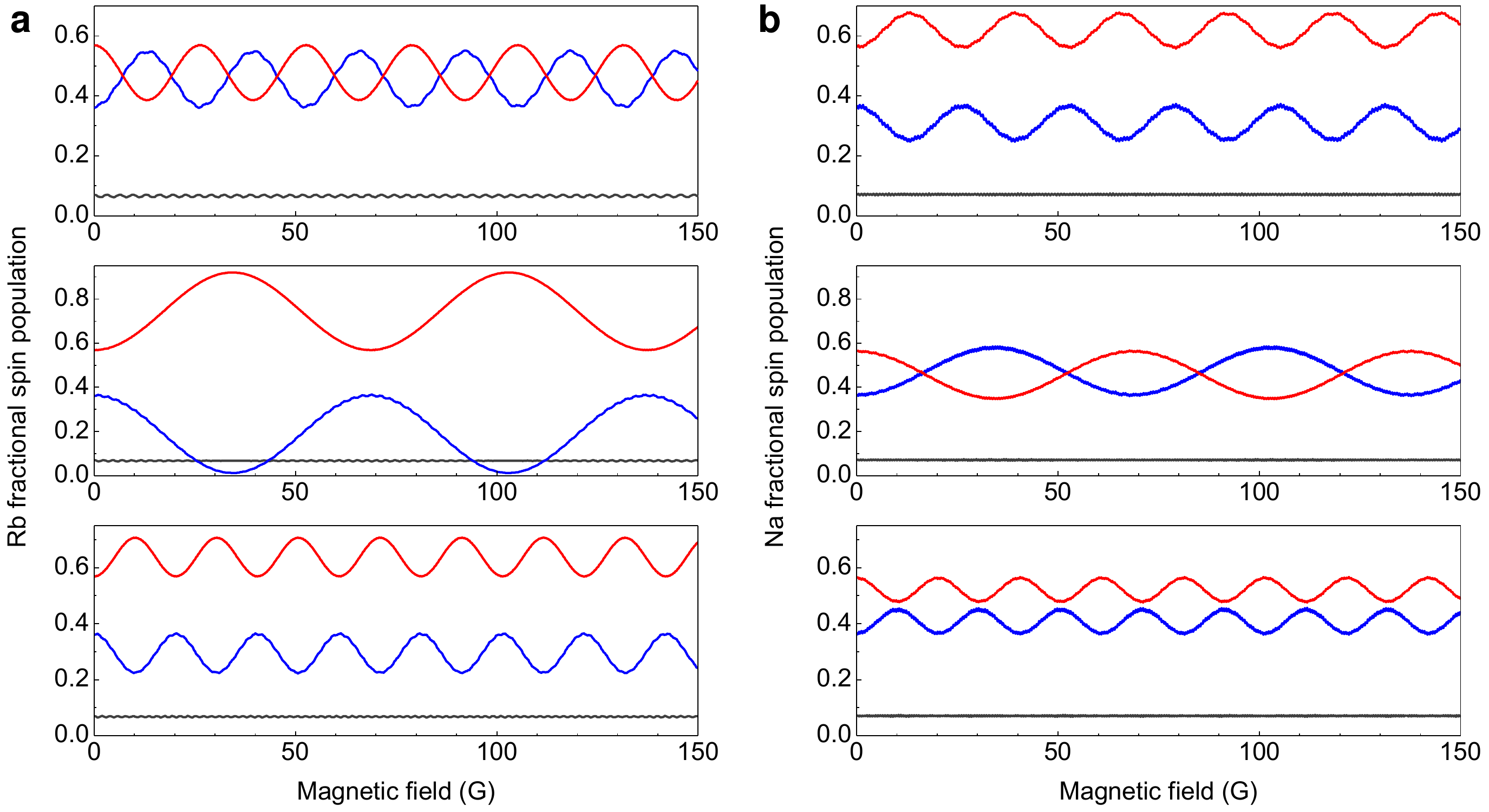}
\caption{\label{sfig1} Simulated heteronuclear spin oscillations for Rb ({\bf a}) and Na ({\bf b}) based on our theoretical model at three different external magnetic fields $B=1.55$ G (top), $1.7$ G (middle) and $1.8$ G (bottom). The $\pi$-phase differences between oscillations of $\ket{-1}$ (red curve) and $\ket{0}$ (blue curve) states of the same species and the same spin state of different species, are well reproduced in our simulations. }
\end{figure}

In this subsection, we discuss in some more detail the numerical calculations of the spin dynamics in a mixture of spinor Bose gases, based on the many-body theory outlined in Sec.\ref{mbt}. We shall focus on two aspects of the spin dynamics in this section: (1) the period, amplitude and phase of the oscillation and (2) the double peak feature of spin resonance.

In Fig.\ref{sfig1}, we present the numerical simulations for three different magnetic fields $B=1.55$G, $1.7$G and $1.8$G. To the left of the resonance, $B=1.55$ G, the oscillation is far off resonance, which results in large frequency oscillation with, however, smaller amplitude. We also note that the phase of the oscillation are consistent with the experimental observation. In Fig.\ref{sfig1}, we also show that the populations in the remaining states $m_{\rm F}=1$ stay almost a constant throughout the simulation. Similarly, for $B=1.8$ G, to the right of the resonance, the oscillation frequency is high while the amplitude is small. The initial oscillation phase of the $m_{\rm F}=0$ is opposite to that at $B=1.55$ G and is consistent with experiment. Close to resonance at $B=1.70$ G, the amplitude of the oscillation is very large while the frequency is very low. All the above are qualitatively consistent with two-particle picture.

\begin{figure}
\includegraphics[width=0.6 \linewidth]{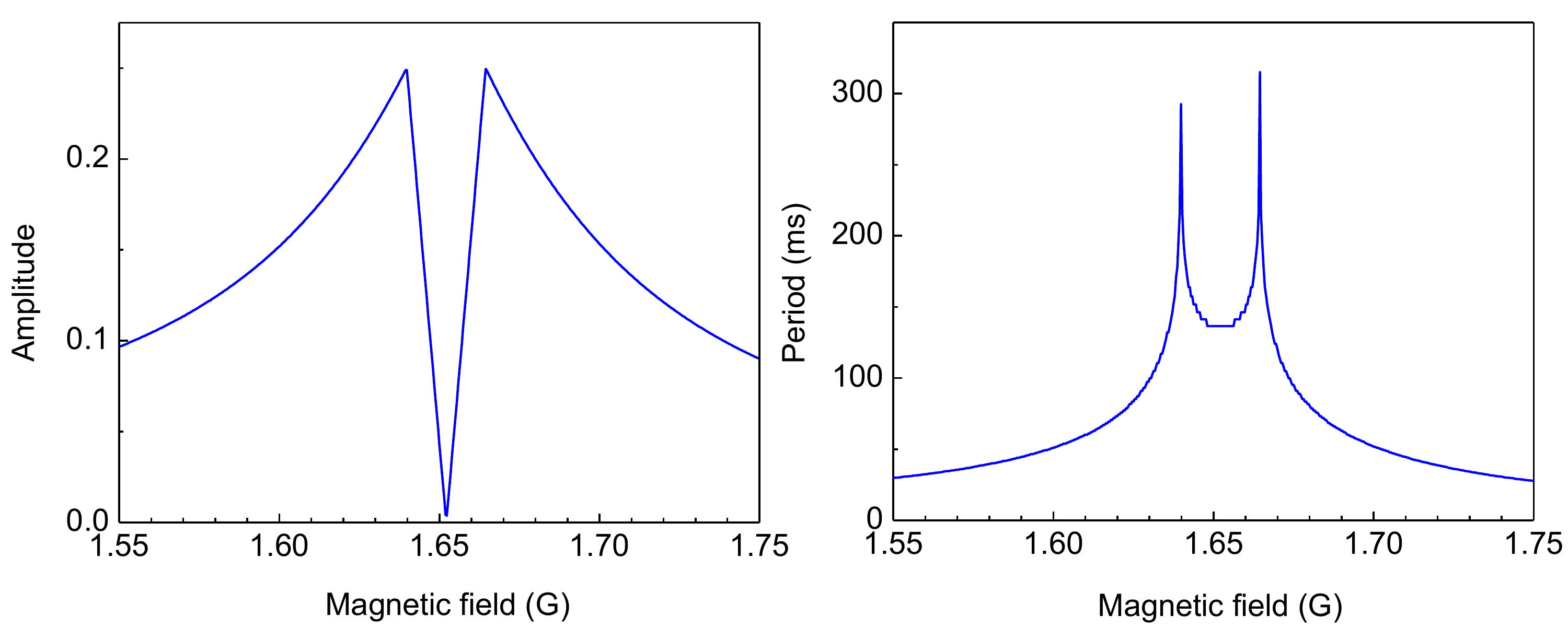}
\caption{\label{sfig2} Simulated spin oscillation amplitudes ({\bf a}) and periods ({\bf b}) with the magnetic field near $B_0$ with equal numbers of Na and Rb atoms. Na and Rb have the same oscillation period and amplitude here. Double peaks are clearly shown in this case.}
\end{figure}

A qualitatively new feature emerges close to the resonance, where the single resonance peak predicted by the two-body theory split into two. This feature depends on the initial state of the spinor mixture and furthermore, depends on the relative population of the two species. In Fig.\ref{sfig2}, we show the amplitude and frequency of the spin mixing for an equal mixture of Rb and Na atoms. Clear double peak structure are observable in both amplitude and frequency. While in our experiment, the double peak is only apparent in the amplitude due to the number imbalance.

Finally, in our numerically simulation, the effects of damping have not been included. However, one feature that is common to our  experimental observation is that the damping has the strongest effect close to resonance. This can be explained by noticing that the effects of spin-spin interaction is the strongest, as compared with the single particle Zeeman energy in this region. As a result, one expects significant damping there. Other sources of damping may come from the residual magnetic field gradient in our experiment, which tends to drive the sample into incoherent mixture through spin diffusion. This later effect is very important in our system also since the Na condensate cloud is much larger than the Rb cloud and spin density gradient are very large especially close to the interface of the two cloud. We intend to study these effects more carefully in a separate publication.